\documentclass[sigconf]{acmart}
\AtBeginDocument{%
  }

\copyrightyear{2026}
\acmYear{2026}
\setcopyright{cc}
\setcctype{by}
\acmConference[CHI EA '26]{Extended Abstracts of the 2026 CHI Conference on Human Factors in Computing Systems}{April 13--17, 2026}{Barcelona, Spain}
\acmBooktitle{Extended Abstracts of the 2026 CHI Conference on Human Factors in Computing Systems (CHI EA '26), April 13--17, 2026, Barcelona, Spain}
\acmDOI{10.1145/3772363.3798806}
\acmISBN{979-8-4007-2281-3/2026/04}

\begin{document}

%%
%% The "title" command has an optional parameter,
%% allowing the author to define a "short title" to be used in page headers.
\title{Smells Like Fire: Exploring the Impact of Olfactory Cues in VR Wildfire Evacuation Training}

%%
%% The "author" command and its associated commands are used to define
%% the authors and their affiliations.
%% Of note is the shared affiliation of the first two authors, and the
%% "authornote" and "authornotemark" commands
%% used to denote shared contribution to the research.
\author{Alison Crosby}
\email{arcrosby@ucsc.edu}
\orcid{0000-0002-3610-1319}
\affiliation{%
  \institution{University of California, Santa Cruz}
  \city{Santa Cruz}
  \state{California}
  \country{USA}
}

\author{MJ Johns}
\email{mljohns@ucsc.edu}
\orcid{0009-0002-4016-2517}
\affiliation{%
  \institution{University of California, Santa Cruz}
  \city{Santa Cruz}
  \state{California}
  \country{USA}
}

\author{Eunsol Choi}
\email{echoi33@ucsc.edu}
\orcid{0000-0002-4397-5751}
\affiliation{%
  \institution{University of California, Santa Cruz}
  \city{Santa Cruz}
  \state{California}
  \country{USA}
}

\author{Tejas Polu}
\email{tpolu@ucsc.edu}
\orcid{0009-0008-3972-7342}
\affiliation{%
  \institution{University of California, Santa Cruz}
  \city{Santa Cruz}
  \state{California}
  \country{USA}
}

\author{Katherine Isbister}
\email{katherine.isbister@ucsc.edu}
\orcid{0000-0003-2459-4045}
\affiliation{%
  \institution{University of California, Santa Cruz}
  \city{Santa Cruz}
  \state{California}
  \country{USA}
}

\author{Sri Kurniawan}
\email{skurnia@ucsc.edu}
\orcid{0000-0003-0372-5800}
\affiliation{%
  \institution{University of California, Santa Cruz}
  \city{Santa Cruz}
  \state{California}
  \country{USA}
}

%%
%% By default, the full list of authors will be used in the page
%% headers. Often, this list is too long, and will overlap
%% other information printed in the page headers. This command allows
%% the author to define a more concise list
%% of authors' names for this purpose.
\renewcommand{\shortauthors}{Crosby et al.}

%%
%% The abstract is a short summary of the work to be presented in the
%% article.
\begin{abstract}
This paper presents a pilot study exploring the effects of an olfactory stimulus (smoke) for a Virtual Reality game designed to support wildfire evacuation preparedness. Participants (N=18) were split evenly into either a smoke or a control condition, and both completed the same evacuation task. Post-task surveys assessed the participants' perceived preparedness and overall experience.
Initial findings suggest participants in the smoke condition reported significantly higher immersion compared to those in the control condition. Across both groups, participants expressed an increased sense of preparedness for real-world wildfire evacuations following the experience. %Overall, our research indicates that incorporating olfactory cues may enhance the immersiveness of VR-based training tools for emergency preparedness.
\end{abstract}

%%
%% The code below is generated by the tool at http://dl.acm.org/ccs.cfm.
%% Please copy and paste the code instead of the example below.
%%

%%
%% Keywords. The author(s) should pick words that accurately describe
%% the work being presented. Separate the keywords with commas.
%% A "teaser" image appears between the author and affiliation
%% information and the body of the document, and typically spans the
%% page.
\begin{teaserfigure}
  \includegraphics[width=\textwidth]{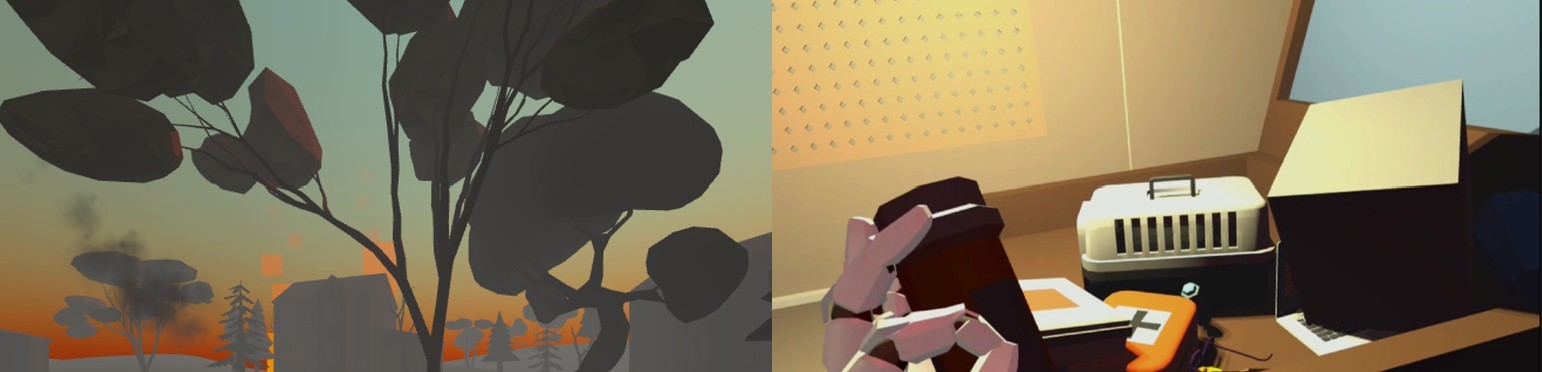}
  \caption{A view of what players can see from a window in the house. A loaded truck filled with items to evacuate with.}
  \Description{The view of smoke and fire from a window in the house. And an example of one participant's packed truck.}
  \label{fig:teaser}
\end{teaserfigure}

%\received{20 February 2007}
%\received[revised]{12 March 2009}
%\received[accepted]{5 June 2009}

%%
%% This command processes the author and affiliation and title
%% information and builds the first part of the formatted document.
\maketitle

\section{Introduction}
Our research explores the use of Virtual Reality (VR) for wildfire evacuation preparation training. The work presented in this paper builds upon prior work that used a VR simulation game to target the pre-evacuation process of packing and preparing before a disaster \cite{crosby_designing_2025}. For our study, we sought to investigate the effects of including an olfactory stimulus, smoke. Smoke is a constant by-product of wildfires and has harmful, long-reaching effects on communities even hundreds of miles away from the source \cite{gould2024health}. People may even be first alerted to a potential wildfire through the smell of smoke in the air. This cue of nearby danger acts as an initial prompt to be aware of one's surroundings and even start the evacuation process. There is also a risk for smoke to trigger anxiety, depression, or post-traumatic stress disorder, especially for those who have previously experienced a wildfire \cite{humphreys2022can, eisenman2022mental}. This may then lead to an increase in evacuation time and inhibited decision-making abilities, which places further strain on emergency responders and resources during a disaster.
To this end, we sought to investigate the use of a smoke scent to create a more immersive VR experience, as well as better prepare users for a wildfire evacuation. 

This paper presents a between-subjects (N = 18) user study comparing the use of an olfactory stimulus (smoke) to a control group with no olfactory stimulus (no smoke). We assess the differences in conditions using usability scales tailored to the experience, and participant-reported feelings of preparedness (both pre- and post-experience). We also collected qualitative data in the form of feedback to assist with understanding the quantitative results and provide guidance for future design.
This pilot study sought to answer: 1) \textit{Do participants in the smoke condition report higher feelings of wildfire preparedness?} and 2) \textit{Do participants in the smoke condition report increased immersion?}
Our findings show significantly higher scores for feelings of preparedness across all participants post-study, suggesting the benefit of, and further support for, the development of disaster-related VR training scenarios. Participants in the smoke condition also perceived the game as more immersive and felt more knowledgeable about what items to pack in a wildfire evacuation scenario. Implications for future research include further development of multimodal approaches to enhance immersion and knowledge acquisition in VR disaster training.

\section{Related Work}

%\subsection{Virtual Reality for Emergency and Evacuation Training}
VR is commonly used for training high-risk procedures because it allows trainees to confront realistic hazards without physical danger. Meta-analyses across domains show robust gains in knowledge, spatial reasoning, and psychomotor skill when training is delivered in VR \cite{Abich2021,Feng2018}. The momentum is no longer limited to safety-critical industries: even computer-science laboratories are adopting head-mounted simulations to scaffold practical skills \cite{Ebrahimi2024}.
As the medium matures, both software pipelines and dedicated peripherals are being refined for emergency contexts. Jarvis et al. chronicle how successive iterations of engines, haptics, and display hardware have increased the fidelity of explosion and fire safety simulations over the past decade \cite{Jarvis2021}. Complementing these technical advances, Blackler et al.\ argue that user-centered interaction design is essential for translating fidelity into actual preparedness gains; their design framework aligns scenario goals, interface affordances, and assessment metrics within a single workflow \cite{Blackler2024}.

Applied research confirms that well-designed simulations can reproduce the pressure of building evacuations; multisensory office fire scenarios prompt realistic route choices and timing behaviors \cite{Shaw2019}, and evolutionary VR models have been used to test school building exit strategies with hundreds of synthetic occupants \cite{Lorusso2022}. These studies underscore the versatility of VR across scenarios for building/ structure fires.
Outdoor wildfire scenarios are beginning to broaden that focus. Vega et al. demonstrated a wildfire-related game by teaching campfire safety in a gamified forest environment \cite{Vega2017}. Formative work by Loh et al. gathered requirements from recent wildfire evacuees, highlighting communication, reflection, and item-management needs that could shape VR scenario design \cite{Loh2023}. More recently, Crosby and Johns introduced a standalone headset experience that asks players to pack possessions and evacuate under an advancing fire front, with early tests suggesting strong memorability and preparedness effects \cite{Crosby2024}. Nevertheless, empirical work on the measurable transfer of evacuation skill remains sparse.

%\subsection{Olfactory Stimuli and Multisensory Immersion in VR}
Olfactory immersion adds a dimension of memory and emotion that purely visual or auditory cues cannot match. Herz argues that olfactory VR can support exposure therapy and help prevent post-traumatic stress by harnessing scent-triggered emotional memories \cite{Herz2021}. Given this, olfaction demonstrated strong potential in learning and education with increased recall and better problem solving \cite{youngblut1996review, washburn2003olfactory}. Beyond clinical contexts, scent is recognized as a powerful empathy tool. Pratte et al.\ propose a framework in which multisensory artifacts, including smell, deepen users’ emotional engagement with another person’s experience \cite{Pratte2021}. Engaging the public with real smoke exposure, Rappold et al.\ show that citizen-science participants who self-report wildfire smoke become more aware of health risks and adopt protective actions \cite{Rappold2019}. These studies suggest that scent influences users’ emotions and behavior to enhance learning and engagement.

Smell also alters spatial perception and presence. Vilaplana and Yamanaka found that adding ambient lavender or orange scent changed how participants judged the size and pleasantness of a waiting room \cite{Vilaplana2015}. In VR specifically, ambient odors can heighten realism and help reduce simulator sickness when the scent matches the virtual context \cite{Ranasinghe2020}. Spatial congruence is critical, however. Tsai et al.\ showed that when visual cues were misaligned with an odor source, users relied on sight and mislocated the scent, which reduced their sense of directionality \cite{Tsai2021}. Overall, the literature indicates that olfactory cues can boost presence, empathy, and behavioral realism in VR. These insights motivate our exploration of simulated wildfire smoke as an integral component of evacuation training.

\section{Method}
This work evaluates a between-subjects user study with 18 participants and two conditions. The two conditions compared the inclusion of an olfactory stimulus, with one condition including a smoke-scent while participants played the VR game, while the control condition had no added scent. 
This study received IRB approval to test on a university campus with both student and non-student participants.
The study took place at two separate locations associated with the university: one in a wildfire-risk city and the other in a not-at-risk city. For consistency, the smoke-scent condition was conducted in the same lab in the location not at risk of wildfires. The VR simulation game was run on a Meta Quest 2 headset.

Participants were recruited from a university campus, including both undergraduate and graduate students, through flyers and outreach emails to specific student groups and classes. In total, our sample size consisted of 18 participants (7 Female, 7 Male, 2 Non-binary / third gender, and 2 preferred not to report), with ages ranging from 20 to 33 years old (\textit{M} = 26 years, \textit{SD} = 3.4 years). We recruited participants who had never evacuated from a wildfire before, due to the risk of triggering unpleasant memories and causing unintentional harm.
The majority of participants (12) reported having at least ``a little'' experience with virtual reality; no participant reported having no experience. 
Participants were told they could end the task at any point, especially if they were experiencing symptoms of motion sickness. One participant took a water break between the tutorial and the main task; Every participant was able to complete the task.

Participants signed an informed consent form and completed a pre-survey that included demographic questions and questions about current levels of preparedness. Participants were then introduced to the VR experience by a researcher and spent time in the tutorial scene, becoming familiar with the controls. Once ready, the timed evacuation experience began, in which the participants were tasked with navigating a virtual house to pack items. Rather than being given a specific list of items to collect, participants were to consider what they felt would be important- including both critical items like pets and supplies and sentimental items like photo albums. Upon completing the evacuation, participants were presented with a list of the items they packed as well as recommended items listed on the Cal Fire's website\footnote{\url{https://readyforwildfire.org/prepare-for-wildfire/wildfire-action-plan/}}). Participants then completed post-surveys, including a system usability survey and questions about preparedness. 

\subsection{System Design}
The VR experience, provided by \cite{crosby_designing_2025}, simulates an evacuation packing experience that allows players to move between rooms, pick up items, and load things into a truck or packing boxes spread around the house. It was designed in Unity and built to run on the Meta Quest 2. Participants can use the joysticks or a teleportation option to move around. The triggers on the controllers were used to pick up items. %and must be physically close to the virtual item, requiring them to reach up or crouch down to get some items. 
This experience leverages and builds on past preliminary work from the designers, such as interviews with wildfire evacuees about their experience preparing to leave \cite{Loh2023}, a prototype of a VR packing game \cite{Crosby2024}, and a pilot study exploring how users experience the stress of an evacuation in VR \cite{crosby_designing_2025}.

For the smoke scent, an essential oil diffuser\footnote{Brookstone Ambient Flame Ultrasonic Diffuser} was used. Four droplets (0.2g) each of two different oils, Campfire\footnote{From P \& G Trading} and Smoke\footnote{From Good Essential Oil}, were mixed to create the smell.
The game did not incorporate event-synchronous scent diffusion or localized olfactory stimuli; the scent diffuser operated in the background as ambient smell, much like the in-game interaction and real wildfire scenarios. The scent diffusion was turned on at the beginning of the task (after the tutorial scene) until the participant finished the game. 
For those in the smoke condition, the scent diffuser was set up on a desk just outside the VR play area, approximately 4 feet from the center of the play area. Participants were not informed about the smoke scent before beginning the test or while they were completing the post-task surveys, so as not to sway their opinions. To ensure no cross-contamination of the smoke scent between sessions, we waited an hour after performing another study and utilized an air purifying machine. %, and used separate room locations in the study space when necessary. 

\section{Results}
 The distribution of survey responses was assessed using Shapiro-Wilk tests because of the sample size (N=18). Some variables showed deviations from normality, so nonparametric analyses were performed: Wilcoxon signed-rank tests to compare pre- and post-survey responses, and Mann-Whitney U tests to compare effects across condition types. 

%\subsection{VR Sickness Questionnaire}
%In a post-task survey, immediately after completing the task in VR, participants respond to a Virtual Reality Sickness Questionnaire (VRSQ) \cite{kim2018virtual}. This scale had participants respond to nine potential symptom types (e.g., headache, fatigue, eye strain) on a 4-point scale ranging from None (0) to Severe (3) to measure their overall sickness. We were interested in testing VR sickness for the overall gameplay experience, but also particularly interested in whether the addition of a smoke smell would have an effect. 
%Following the calculation guidance described in the paper, the mean average scores across all participants were reported as (\textit{M} = 22.64, \textit{Mdn} = 18.75, \textit{SD} = 15.12). 
%The smoke condition reported a lower mean score (\textit{M} = 21.30, \textit{Mdn} = 21.67, \textit{SD} = 14.69) compared to the non-smoke condition (\textit{M} = 23.98, \textit{Mdn} = 15.83, \textit{SD} = 16.31).
%When examining the qualitative feedback portion of the survey, seven participants wrote about feelings of motion sickness; four of the participants were in the smoke condition. Only one participant used the teleportation movement throughout the whole experience. 

%No statistically significant effect was found when performing a Welch's two-sample t-test on condition type. The results include \textit{t}(15.83) = -0.37, \textit{p} = 0.72. The 95\% confidence interval for the mean difference ranged from -18.21 to 12.84.

\subsection{System Usability and Gameplay Experience}

To evaluate the overall usability of the VR game and controls, as well as their experience, participants responded to twelve questions. A full list of system usability and gameplay questions asked can be found at Appendix \ref{sys_us_q}. Participants responded to each question on a scale of Bad (1) to Excellent (5). These questions were created to align with specific aspects of the system design, such as movement, visual clarity, comfort, and ability to accomplish the task. The average scores of each question were assessed and then compared by condition type. Significant results can be found in Table \ref{tab:mann}. 

A Mann-Whitney U test was conducted to compare reported immersion levels between participants who experienced the VR game with the smoke scent and those who did not. Participants responded to the following question: ``How would you rate your immersion experience?''
The mean immersion rating for the smoke condition was higher (\textit{M} = 4.33, \textit{Mdn} = 4, \textit{SD} = 0.71) compared to the control condition (\textit{M} = 3.44, \textit{Mdn} = 4, \textit{SD} = 0.73), indicating a perceived increase in immersion when the olfactory stimulus was present. A statistically significant difference in immersion scores was revealed between the two groups: \textit{W} = 64.5, \textit{p-value} = 0.025. These results indicate that the addition of the olfactory stimulus enhanced participants' sense of immersion.

A statistically significant difference in scores was also found in response to: ``How would you rate your ability to see clearly in the virtual environment?''
Participants in the smoke condition reported higher visual clarity (\textit{M} = 4.00, \textit{Mdn} = 4, \textit{SD} = 0.71) than those in the control condition (\textit{M} = 3.33, \textit{Mdn} = 3, \textit{SD} = 0.50). The significant difference between the two conditions is reported as \textit{W} = 61.5, \textit{p-value} = 0.046. 
To note, two different headsets for this study were used. Statistical analysis was performed comparing the responses to this question with headset type and no significance was found, suggesting that the difference in response was due to the olfactory stimulus.
%A potential explanation for this finding is that the addition of a smoke scent enhanced participants’ perception of environmental cues or their understanding of the virtual scenario. 

% This is the table for Usability and Overall UNPAIRED test results---
%moved to appendix

\begin{table}[h!]
%\begin{threeparttable}
\caption{Mann-Whitney U Tests for Smoke and Non-Smoke Groups}
\label{tab:mann}

\begin{tabular}{p{1.25cm}|p{0.2cm}|p{0.8cm}|p{0.8cm}|p{0.8cm}|p{0.8cm}|p{0.6cm}|p{0.6cm}}

\hline
& N & Smoke Mean & Smoke SD & No Smoke Mean & No Smoke SD & W & $\rho< 0.05$ \\
\hline

Immersion
& 9 & 4.33 & 0.71 & 3.44 & 0.73 & 64.5 & \textbf{0.025}\\

\hline

Visual Clarity
& 9 & 4.00 & 0.71 & 3.33 & 0.50 & 61.5 & \textbf{0.046}\\

\hline

\end{tabular}

%\end{threeparttable}
\end{table}

%Additionally, participants were asked to report their level of stress while completing the VR task, from a range of No Stress (1) to Extreme Stress (7). We performed a Mann-Whitney U test, but found no statistical difference in condition type for this question (\textit{W} = 54, \textit{p-value} = 0.218). However, participants in the smoke condition reported a higher average score (\textit{M} = 3.44, \textit{Mdn} = 3, \textit{SD} = 1.24), compared to the control condition (\textit{M} = 3.00, \textit{Mdn} = 3, \textit{SD} = 1.22).  %These results indicate that the VR game is relatively low in stress. This is an aspect of the game the researchers intend to explore more in future iterations, as feedback from previous user testing has shown some participants' desire for a more stressful scenario. Allowing players to set their own stress levels is likely the best scenario and can also provide more opportunities for repeat playability. 

\subsection{Feeling More Prepared Post-Task}

To evaluate change in feelings of preparedness, knowledge of items to pack, and confidence in locating items, pre-test and post-test Likert responses for the following statements were compared: 1) ``I feel prepared for a wildfire evacuation.'', 2) ``I know what items to pack for a wildfire evacuation.'', and 3) ``I can easily locate all necessary items for an evacuation around my house.''
The Likert values (Strongly Disagree to Strongly Agree) were converted to a 7-point scale, and scores were averaged across participants within both conditions. In the pre-test, the average for feelings of preparedness was 3.06 (\textit{SD} = 1.30), which increased to 5.33 (\textit{SD} = 0.97) in the post-test. For knowledge of items to pack, the average score increased from 3.61 (\textit{SD} = 1.65) in the pre-test to 5.83 (\textit{SD} = 1.15) in the post-test. Confidence in finding items increased from an average of 3.83 (\textit{SD} = 1.92) in the pre-test to 5.50 (\textit{SD} = 0.79) in the post-test. These results can be seen below in Figure \ref{fig:stats}. 

\begin{figure}[h]
    \includegraphics[width=.65\linewidth]{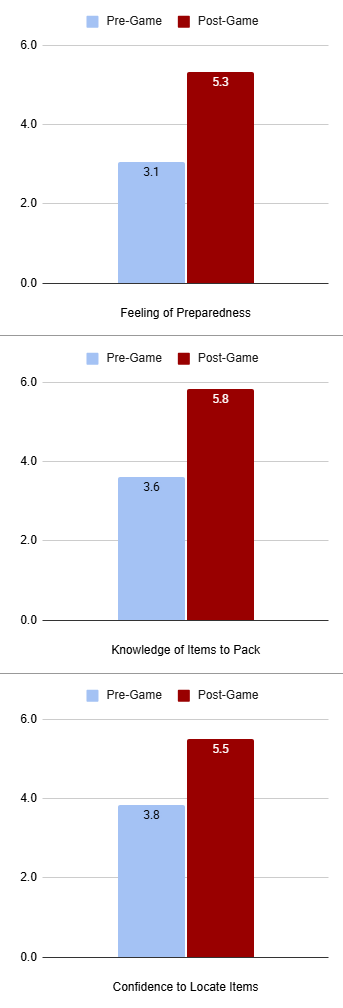}
    \Description{Comparing the pre-test and post-test Likert scores for feeling of preparedness, knowledge of items to pack, and confidence in finding items.}
    \caption{Comparing the pre-test and post-test Likert scores for feeling of preparedness, knowledge of items to pack, and confidence in finding items.}
    \label{fig:stats}
\end{figure}

A  Wilcoxon test was conducted to compare pre- and post-test scores across all participant data for the three statements listed above. All three questions reported a statistically significant ($\rho< 0.05$) difference in scores. These findings indicate that participants felt more prepared for a wildfire evacuation after playing the VR game. See Table \ref{tab:wilcoxon} for an overview of the findings. 

% This is the table for PAIRED wildfire preparedness results---
\begin{table}[ht]
\caption{Wilcoxon Signed-Rank Tests for Wildfire Preparedness Survey Questions}
\label{tab:wilcoxon}

\begin{tabular}{p{2cm}|p{0.4cm}|p{0.7cm}|p{0.7cm}|p{0.7cm}|p{0.8cm}|p{1cm}}

\hline
 & N & Pre- Mean & Post- Mean & W+ (V) & $\rho< 0.05$ \\
\hline

Feeling More Prepared
& 18 & 3.06 & 5.33 & 7.5 & \textbf{0.001}\\

\hline

Know What to Pack
& 18 & 3.61 & 5.83 & 0 & \textbf{0.001}\\

\hline

Location of Evacuation Items
& 18 & 3.83 & 5.5 & 3 & \textbf{0.005}\\

\hline

\end{tabular}
\end{table}

When asked whether the game promoted any potential behavior change, nine participants responded ``yes'', eight responded ``unsure'', and one responded ``no''. For the nine participants who responded ``yes'', they were then prompted with what type of behavior change they would engage in. Six responded that they would attempt to prepare a go-bag. Additionally, four participants mentioned being more conscious and aware of the items they would want to pack, as well as where these items may be located in their house. P3 wrote that the game ``was a great opportunity for me to revisit on the idea of how we should be always prepared for the worst. This VR game was a good reminder. I would definitely go home and prepare my evacuation bag (Go-bag).''

% This is the table for PAIRED wildfire preparedness results---
%moved to appendix

\subsubsection{Comparing Conditions Post-task}

A Mann-Whitney U test was performed on the same statements for the post-test results to compare the two conditions and assess any statistical difference in scores based on the presence of the smoke scent. 
Statistical significance  ($\rho< 0.05$) was only found in response to the second statement. Compared to the non-smoke condition, participants in the smoke condition felt more knowledgeable about what items to pack during a wildfire evacuation post-test; \textit{W} = 70.5, \textit{p-value} = 0.007. 
This result is supported by qualitative feedback given as part of the post-study survey. P13 stated that they ``really liked having to collect items from around the house and it was helpful to see the list of important items at the end of the game. It felt like it really replicated the experience of being panicked and trying to pack and it made me feel like I would be better equipped to handle that situation in real life.''
On average, participants in the smoke condition reported higher scores post-task, as seen in Figure \ref{fig:conditionstats} within the appendix.

%"I feel prepared for a wildfire evacuation.": t = 2.1381, df = 13.517, p-value = 0.0513

%"I know what items to pack for a wildfire evacuation.": t = 3.3849, df = 11.135, p-value = 0.00599

%"I can easily locate all necessary items for an evacuation around my house.": t = 1.5617, df = 15.264, p-value = 0.1388

\subsection{Overall Experience}

Mann-Whitney U tests were conducted to compare participant responses between the smoke and no-smoke VR conditions across four experience-related questions, see questions 1-4 in Appendix \ref{wp_q}. Overall, participants in the smoke condition reported significantly higher scores on all items (\textit{p-value} \textless 0.05), which suggests enhanced perceived effectiveness and realism of the game experience when olfactory cues were present. %The statements include:
%\begin{enumerate}
%    \item ``I felt like I successfully evacuated with every item I needed.''
%    \item ``I felt like I was in a real-life evacuation experience.''
%    \item ``I feel more prepared to evacuate from a wildfire after this VR experience.''
%    \item ``I would complete this VR experience again.''
%\end{enumerate}

%\textbf{``I felt like I successfully evacuated with every item I needed.''} -- 
Participants in the smoke condition (\textit{M} = 6.44, \textit{Mdn} = 7, \textit{SD} = 0.88) reported feeling more successful in evacuating with all necessary items compared to those in the control condition (\textit{M} = 5.33, \textit{Mdn} = 6, \textit{SD} = 1.22); the Mann-Whitney U test reported \textit{W} = 63, \textit{p-value} = 0.042. 
%\textbf{``I felt like I was in a real-life evacuation experience.''} -- 
Participants in the smoke condition (\textit{M} = 5.44, \textit{Mdn} = 6, \textit{SD} = 1.13) were significantly more likely to report that the VR experience felt more realistic than those in the control condition (\textit{M} = 3.89, \textit{Mdn} = 4, \textit{SD} = 1.27), \textit{t}(15.79) = 2.75; the Mann-Whitney U test reported \textit{W} = 66.5, \textit{p-value} = 0.021. 
%\textbf{``I feel more prepared to evacuate from a wildfire after this VR experience.''} -- 
Participants in the smoke condition (\textit{M} = 6.44, \textit{Mdn} = 7, \textit{SD} = 0.73) reported feeling more prepared to evacuate from a wildfire after the VR experience compared to the control condition (\textit{M} = 5.22, \textit{Mdn} = 6, \textit{SD} = 1.30); the Mann-Whitney U test reported \textit{W} = 64, \textit{p-value} = 0.032. 
%\textbf{``I would complete this VR experience again.''} -- 
Participants in the smoke condition (\textit{M} = 6.11, \textit{Mdn} = 6, \textit{SD} = 0.78) were more likely to say they would repeat the VR experience than those in the control condition (\textit{M} = 5.33, \textit{Mdn} = 5, \textit{SD} = 0.50); the Mann-Whitney U test reported \textit{W} = 63, \textit{p-value} = 0.035. 

%Besides the smoke scent, both study conditions employed audio cues to increase immersion and urgency. Based on qualitative feedback from participants P3, P4, P5, P6, and P8, the seriousness of the sirens that can be heard moving along the street throughout the game, the narrator's alerts, and the fading alarms of firefighter trucks reinforced the feeling of time slipping away. The physical clock with its ticking animation increased the pressure, prompting one participant (P6) to say it was ``quite stressful that made me rush things.'' Visual cues that there was a fire nearby included fire particles off in the distance, behind a house, and smoke particles outside a few windows. The red glow of the sky also assisted in notifying participants that a fire was nearby.
 
\section{Discussion}

%We investigated the inclusion of an olfactory stimulus (smoke) for a wildfire evacuation simulation game. In our study, we compared a smoke-scent condition group to a control group (no-smoke scent). Our findings suggest that the inclusion of the scent increased immersion. Participants in the smoke condition also reported feeling more knowledgeable about items to pack in an evacuation scenario, indicating a perception that they could see more clearly in the virtual environment. Overall, participants in both conditions reported feeling more prepared for a wildfire evacuation after playing. In the following subsections, we discuss insights, suggest future work, and note the limitations of this study. 

%\subsection{Increased Preparedness for a Wildfire}
The quantitative results comparing pre-test and post-test responses show an increase in feelings of preparedness, knowledge of what items to pack, and confidence in locating items, suggesting that the VR game is effective at helping participants learn and practice the pre-evacuation skill of packing. Existing literature indicates that opportunities to practice pre-evacuation behavior may help reduce the negative effects of stress and anxiety on decision-making skills and cognitive performance \cite{mclennan2011bushfire}. There is real value in enabling participants to have a "first-hand" experience in VR to practice the steps of what items might actually be important in an evacuation scenario- even with the house and items not exactly replicating a person's true experience.

The results of our limited pilot study suggest that the inclusion of the olfactory stimulus (smoke) increased immersion and positively influenced participants’ perceived effectiveness, realism, and engagement with the VR-based wildfire evacuation training game.
The immersion aspect is both important and impactful for training simulations, as it allows users to better connect to a high-risk, high-stress scenario like a wildfire evacuation. Increased immersion may also increase a user's ability to recall their experience for future benefit \cite{mania2001effects}, suggesting potential for a subsequent study to examine long-term effects. VR has shown great promise and effectiveness for empathy boosting by allowing users to view new perspectives through its inherent immersive properties that enable users to feel present in the moment \cite{ventura2020virtual, shin2018empathy}. This empathy boosting may allow users who do not live in wildfire zones to understand the risk and challenges other communities face, which could boost support and relief measures in times of crisis. 
This is also supported by the statistically significant finding in response to the question about the visual clarity of the virtual environment. The addition of the smoke scent potentially enhanced participants’ perception of environmental cues or their understanding of the virtual scenario. %Though perhaps this question needs restructuring to better address the task, e.g. ``How would you rate your ability to navigate in the virtual environment?'' %An additional explanation may be that their eyes/ brain may have expected a lack of visual clarity because of the smoky scent in the air. However, there was no actual haze fogging their view in the game, resulting in a higher perceived visual acuity.

%While players found several aspects of the game to feel realistic and support preparedness through practice, some pointed out areas where more realism is needed to replicate how they expect the experience would feel in real life. For example, P10 stated that "I sort of wish some of the important items were more hidden or difficult to find -- in real life, I don't currently know exactly where everything is, so mimicking that to some extent might be helpful."
%Additionally, P11 noted that the ``House organization was not in line with how mine is (understandably), but because of that it didn't feel like I was getting to prepared.'' A design aspect that could be investigated further is allowing players to choose from different house types that more accurately depict their real-life situations, such as a multi-family home, roommates, or other pets. The incorporation of different lifestyles not only has the potential to connect with more people but also has the opportunity to introduce features of replayability.

A significant limitation of this pilot study is the small sample size (N=18), which may have biased our findings. Additionally, the study was tested primarily on young adults within a university setting who were familiar with VR. 
Apart from the participant pool, the study took place in two different university locations located in different cities. Two different headsets were also used, though both were Quest 2 models. Additionally, the diffuser used to emit the smoke scent was quite small and may not have had as great an impact as we originally intended.

\section{Conclusion}

This paper shares the results of a comparative user test of a VR experience for wildfire preparedness training, comparing an olfactory condition (smoke scent) to a control condition (no smoke scent). The study (N=18) split participants into two conditions, with nine participants in each, and employed pre- and post-tests to measure change in feelings of preparedness, knowledge, and immersion. The study found statistically significant improvement across all participants in feelings of preparedness, knowledge of what to pack, and confidence in the ability to pack items. The results between condition types also revealed a statistical significance in the feeling of immersion and visual clarity during the experience for the olfactory condition. These results, along with comments from participants, support creating immersive multimodal VR experiences to increase preparedness for wildfires.

%\begin{acks}
%This material is based upon work supported by the National Science Foundation under Grant No. 2230636. 
%\end{acks}

%%
%% The acknowledgments section is defined using the "acks" environment
%% (and NOT an unnumbered section). This ensures the proper
%% identification of the section in the article metadata, and the
%% consistent spelling of the heading.

%%
%% The next two lines define the bibliography style to be used, and
%% the bibliography file.
\bibliographystyle{ACM-Reference-Format}
\bibliography{sample-base}

%%
%% If your work has an appendix, this is the place to put it.
\appendix

\section{System Usability and Gameplay Questions}
\label{sys_us_q}
Participants rated their agreement with the questions on a 1 to 5 Likert scale; 1(Bad), 2 (Poor), 3 (Fair), 4 (Good), 5 (Excellent). 
\begin{enumerate}
    \item How would you rate your ability to control objects in the virtual environment?
    \item How would you rate your ability to move in the virtual environment?
    \item How would you rate your ability to see clearly in the virtual environment?
    \item How would you rate your comfort using the Oculus controllers?
    \item How would you rate your comfort wearing the Oculus headset?
    \item How would you rate your comfort in the virtual environment?
    \item How would you rate your ability to accomplish the tasks asked of you?
    \item How would you rate your immersion experience?
    \item How would you rate your overall enjoyment?
    \item How would you rate your overall stimulation?
    \item How would you rate the overall practicality of the virtual environment?
    \item How would you rate the overall effectiveness of virtual reality for evacuation training?
\end{enumerate}

\section{Wildfire Preparedness Questions}
\label{wp_q}
Participants rated their agreement with the questions on a 1 to 7 Likert scale. 
\begin{enumerate}
    \item ``I felt like I successfully evacuated with every item I needed.''
    \item ``I felt like I was in a real-life evacuation experience.''
    \item ``I feel more prepared to evacuate from a wildfire after this VR experience.''
    \item ``I would complete this VR experience again.''
    \item ``How likely are you to recommend this VR experience to a friend/ family member?''
    \item ``Did this experience promote a change in behavior for you?''
    \item ``I feel prepared for a wildfire evacuation.''
    \item ``I know what items to pack for a wildfire evacuation.''
    \item ``I can easily locate all necessary items for an evacuation around my house.''
\end{enumerate}

\section{Additional Figures}

\begin{figure}[h]
    \includegraphics[width = 0.71\linewidth]{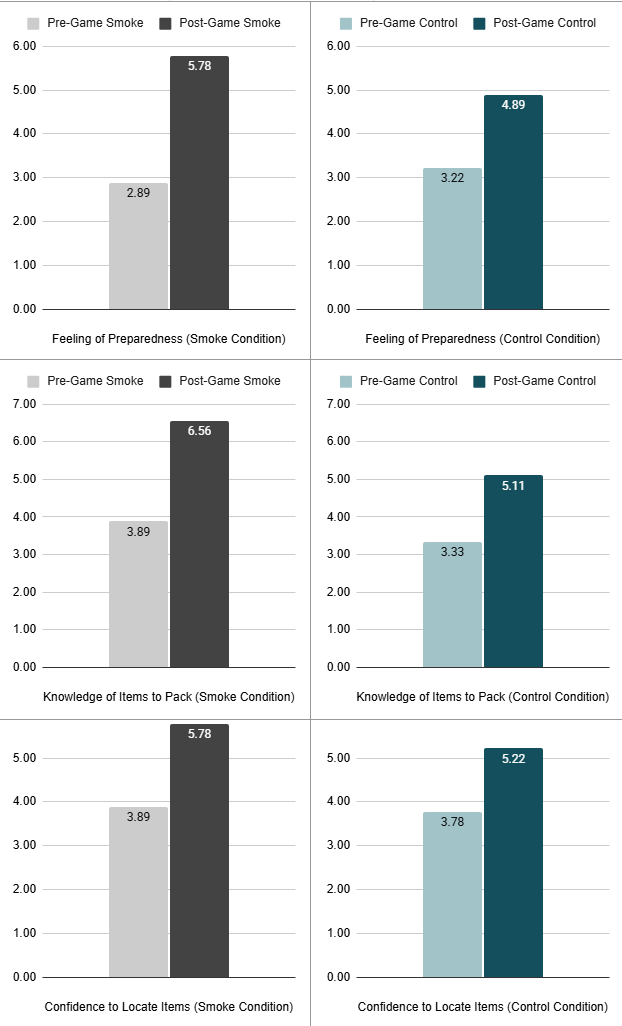}
    \Description{Comparing the pre-test and post-test Likert scores of the smoke condition and control condition (no scent) for feeling of preparedness, knowledge of items to pack, and confidence in finding items.}
    \caption{Comparing the pre-test and post-test Likert scores of the smoke condition and control condition (no scent) for feeling of preparedness, knowledge of items to pack, and confidence in finding items.}
    \label{fig:conditionstats}
\end{figure}

%\section{Tables}

% This is the table for Usability and Overall UNPAIRED test results---

% This is the table for PAIRED wildfire preparedness results---

\end{document}